\documentclass{iau} 
\usepackage{graphicx}

\title[Symp. 315~~Star formation in rotating early-type galaxies] 
{Star formation in early-type galaxies:\\ the role of stellar winds and kinematics}

\author[Silvia Pellegrini, Andrea Negri \& Luca Ciotti]   
{Silvia Pellegrini$^1$, Andrea Negri$^2$ \and Luca Ciotti$^1$}

\affiliation{$^1$Dept. of Physics and Astronomy, University of  Bologna, via Ranzani 1,
I-40127 Bologna, Italy \\ email: {\tt silvia.pellegrini@unibo.it} \\[\affilskip]
$^2$CNRS, UMR 7095, Institut d'Astrophysique de Paris, 98bis bvd Arago, F-75014 Paris, France}

\pubyear{2015}
\volume{315}  
\jname{From interstellar clouds to star-forming galaxies}
\editors{P. Jablonka, F. Van der Tak \& P. Andr\'e, eds.}
\begin{document}

\maketitle

\begin{abstract}
Early-type galaxies (ETGs) host a hot ISM produced mainly by stellar
winds, and heated by Type Ia supernovae and the
thermalization of stellar motions. High resolution 2D
hydrodynamical simulations showed that
ordered rotation in the stellar component results in the formation of a centrifugally supported cold
equatorial disc. In a recent numerical investigation we 
found that subsequent generations of stars are formed in this cold disc;
this process consumes most of the cold gas, leaving at the present epoch cold 
masses comparable to those observed. Most of the new stellar mass
formed a few Gyrs ago, and resides in a disc.
\end{abstract}

\section{Introduction}
High resolution 2D hydrodynamical simulations with the ZEUS-MP2 code
showed that ordered rotation in the stellar
component affects significantly the evolution of the hot ISM in ETGs, and, among other effects, 
results in the formation of a centrifugally supported cold
equatorial disc (Negri et al. 2014, hereafter N14). This disc can be extended ($\sim
0.5 -10$ kpc radius),
and as massive as $10^{10}$M$_{\odot}$ in the biggest ETGs. 
Indeed there is evidence that $\sim 50$\%
of massive ETGs host significant quantities of cold gas (Davis et al.  2011, Serra et al.  2014), often in settled
configurations, sharing the same kinematics of the stars, consistent
with an internal origin. Also, the cold  gas seems to provide material for low level star formation (hereafter SF);
and, in the ATLAS$^{\rm 3D}$ sample, 
molecular gas, SF  and young stellar populations are detected only in fast rotators
  (Sarzi et al.  2013, Davis et al. 2014).  
We then added the possibility for the gas to form stars to the simulations of N14, to explore whether SF can
bring the amount of cold gas in the models
more in agreement with observed values, and whether it can explain
the low-level SF activity currently seen to be ongoing in rotating systems.

\begin{figure}[b]
\vspace*{-0.1 cm}
\begin{center}
\includegraphics[width=3.4in]{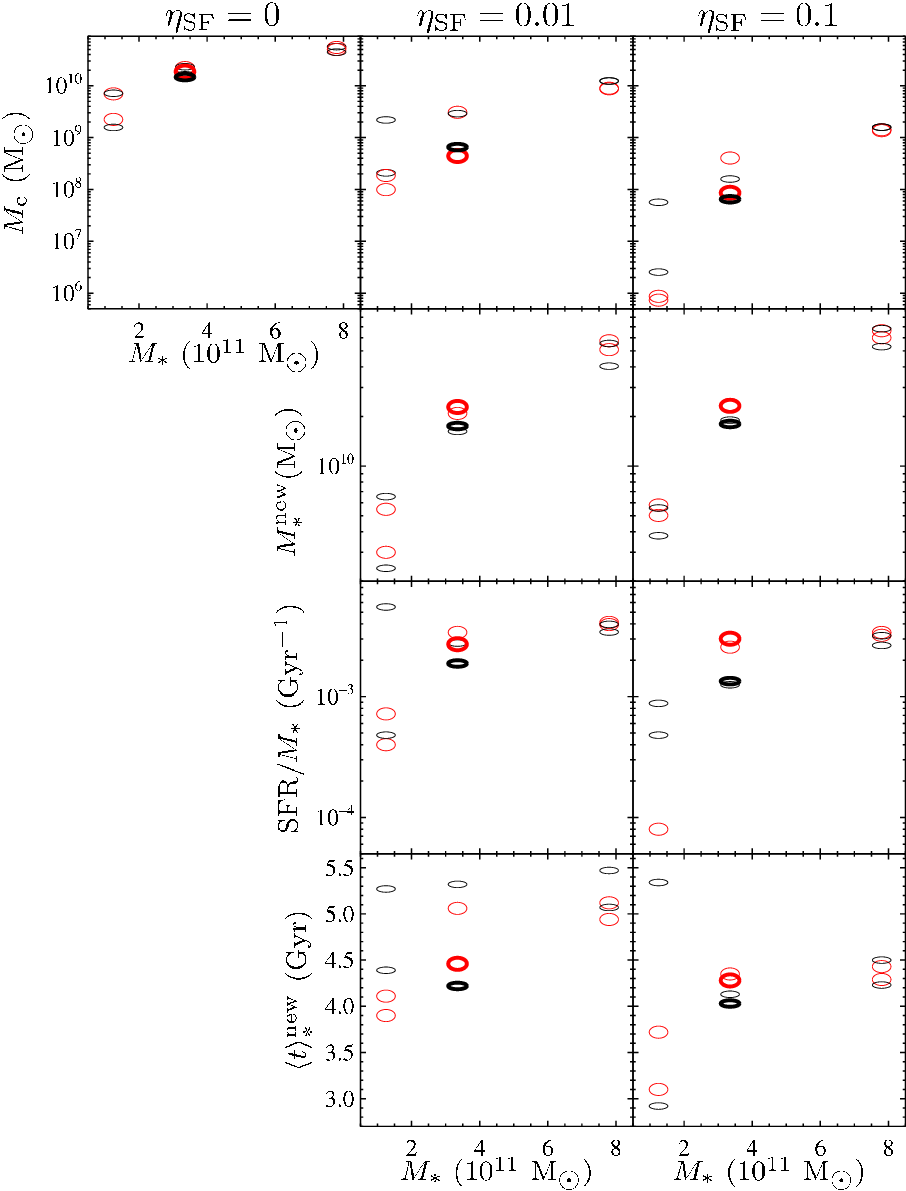} 
\caption{From top to bottom, final values for all rotating models of:
the cold gas mass $M_{\rm c}$ [without SF (left panel), and 
with the two adopted $\eta_{\rm SF}$ values (other panels)];
the stellar mass in newly formed stars $M_*^{\rm new}$; the SF rate
normalized to the initial stellar mass of the galaxy $M_*$;
the mean formation time of the new stars $\langle t\rangle _*^{\rm new}$, 
calculated from an initial time of 2 Gyr.
See Negri et al. (2015) for more details.}
\label{fig1}
\end{center}
\end{figure}

\section{The models and the results}
Hydrodynamical simulations were run for a representative subset of 12 rotating models from N14, including in the code
the removal of cold gas, and the injection of mass, momentum and
energy appropriate for the newly forming stellar population (Negri
et al. 2015).  SF was implemented by
subtracting gas from the grid, at an adopted rate per unit volume of $\dot\rho_{\rm SF} = \eta_{\rm SF} \rho/{t_{\rm SF}}$,
where $\rho$ is the gas density,  $\eta_{\rm SF}$ is the SF efficiency ($\eta_{\rm SF}=0.01$
and $0.1$ were adopted), and ${t_{\rm SF}}$ is the maximum between the cooling timescale 
 and the dynamical timescale. 
In a typical (cyclical) evolution, gas injected by the stellar population
accumulates until radiative losses become catastrophic, 
significant amounts of cold material are produced, and SF is enhanced. At the end 
the cold gas mass $M_{\rm c}$ is much reduced in the models with SF, and the 
mass in new stars $M^{\rm new}_*$ is close to the $M_{\rm c}$ values of the models without SF (Fig. 1).
The new stars reside  mostly in a disc, and could be related to a younger, more metal rich
disky stellar component observed in fast rotators (Cappellari
  et al.  2013).  Most of $M^{\rm new}_*$ formed a few Gyrs ago;
  the SF rate at the present epoch is low ($\le 0.1$M$_{\odot}$yr$^{-1}$), as observed,
  at least for model ETGs of stellar mass $<10^{11}$M$_{\odot}$ (Fig. 1).
The adopted SF recipe reproduces the slope of the
  Kennicutt-Schmidt relation, and even its normalization for $\eta_{\rm SF}=0.01$ (Negri et al. 2015).

\end{document}